# Anisotropic magneto-transport and magnetic properties of low temperature phase of TaTe$_2$

H. X. Chen,[1] Z. L. Li,[1] L.W. Guo,[1*] and X. L. Chen[1,2,3*]

[1] *Research & Development Center for Functional Crystals, Beijing National Laboratory for Condensed Matter Physics, Institute of Physics, Chinese Academy of Sciences, Beijing 100190, China*

[2] *School of Physical Sciences, University of Chinese Academy of Sciences, Beijing 101408, China*

[3] *Collaborative Innovation Center of Quantum Matter, Beijing 100084, China*

**Abstract:** TaTe$_2$ is a quasi-2D charge density wave (CDW) compound with distorted-1T type structure exhibiting double-zigzag chains. Here we report the Fermi surface topology of low temperature phase of TaTe$_2$ (LT-TaTe$_2$) by anisotropic magneto-transport and magnetic measurements on high-quality single crystals. An anomalous large linear magnetoresistance up to 140% at 3 K in 9 T was observed, suggesting the existence of a small Fermi pocket in Dirac cone state in quantum transport models. Meanwhile, strong magnetic anisotropy was observed for $B\perp(001)$ and $B\parallel(001)$. Angle-dependent magnetoresistance and de Hass-van Alphen oscillations suggest the anisotropy of the normal Fermi surface and the small Fermi pocket in Dirac cone state.

---

* Corresponding authors: lwguo@ iphy.ac.cn and chenx29@iphy.ac.cn

# I. INTRODUCTION

Charge density wave (CDW) is a condensed state with the coherent superposition of electron-hole pairs[1], which is usually found in reduced dimensional materials. CDW in quasi-1D systems like TTF-TCNQ[2] is mainly driven by Fermi surface nesting (FSN), whilst the origin of charge density wave in quasi-2D CDW[3] systems has not been well understood and is still in debate. For example, different mechanisms[4] like electron phonon coupling, excitonic insulator instability, and electron interactions beyond FSN are thought to be responsible for the observed CDW in 1T-TaS$_2$, 1T-TiSe$_2$, and the charge order in cuprate superconductors respectively.

Tantalum dichalcogenides (TaX$_2$, X= S, Se, Te) are a class of quasi-2D materials presenting CDW in ground states. Compared with TaS$_2$ and TaSe$_2$, TaTe$_2$ has stronger electron phonon coupling[5] and a larger lattice distortion making it not show various polytypes such as 1T, 2H or any other polytypes. Only a monoclinic phase of TaTe$_2$ (1T'-TaTe$_2$)[6] exists at room temperature, where Ta atoms in 1T'-TaTe$_2$ form a double-zigzag chain-like arrangement[6] along b-axis direction which can be viewed as a commensurate phase by the Peierls transition[7] from a hypothetical 1T phase with a single nesting vector a$^*$/3 where a$^*$ is a reciprocal vector. A Peierls transition occurs at $T_{CDW}$=170 K, and the structure of the low temperature phase of TaTe$_2$ (LT-TaTe$_2$) was determined by Sörgel et al.[8]. Transport, magnetic, heat capacity and thermoelectric properties have been studied and abnormal phenomena[5, 8, 9] were observed around $T_{CDW}$.

The band structures and Fermi surface of 1T'-TaTe$_2$ and hypothetical 1T phase have been studied by using first-principle calculations[7, 10, 11] and other angle-resolved photoelectron emission spectroscopy (ARPES)[7]. But the Fermi surface determined by ARPES is diffusive and experimental results concerning band structure and Fermi surface reconstruction of LT-TaTe$_2$ are still lacking. Clarifying the band structure and the Fermi surface topology are important to unravel the origin of CDW. Here, we aim to study the Fermi surface topology based on analysis of experimental results of magneto-transport and magnetic properties of TaTe$_2$ high quality single crystals grown by us. Anisotropic magneto-transport and magnetic properties of LT-TaTe$_2$ were found and systematically studied. Anomalous linear magnetoresistance (MR) was observed in LT-TaTe$_2$ and possible existence of Dirac cone states was discussed. It is inferred that the anisotropic MR is ascribed to the anisotropy of the small Fermi pocket in Dirac cone state. Our results are helpful in understanding the transport and magnetic properties and to unravel the topology of Fermi surface of LT-TaTe$_2$.

# II. EXPERIMENTAL

Single crystals of TaTe$_2$ were grown by an improved chemical vapor transport (CVT) method[12, 13]. Tantalum foil (99.99%), tellurium powders (99.99%) and 15 mg iodine (99.99%) were loaded into an evacuated silica ampoule. Then the ampoule was heated in a two-zone furnace. The two ends were heated with a rate of 2 K/min to 850°C and 750°C, respectively and the temperature gradient in the two ends were kept for 7 days, and then the ampoule was naturally cooled down to room temperature. Some lamellar single crystals in sizes of over half centimeter in at least one dimension were obtained (Fig.1(b) inset). The chemical compositions were determined by energy-dispersive x-ray analysis (EDX). X-ray diffraction (XRD) pattern was collected using a PANalytical X'Pert PRO diffractometer with Cu radiation.

The resistivity measurements were performed on a Quantum Design physical property measurement system and magnetic properties were measured with vibrating sample magnetometer option with magnetic field parallel or perpendicular the sheet plane of the samples. Due to the crystal growth habits, the monoclinic transition metal dichalcogenides as WTe$_2$, MoTe$_2$, and TaTe$_2$[14] tend to grow as long ribbons along the direction of metal atoms chains. We chose the natural long ribbons of TaTe$_2$ as shown in the inset of Fig.1(a) to measure the in-plane resistivity and MR defined as $(\rho_{xx}(B,T)-\rho_{xx}(0,T))/\rho_{xx}(0,T)$ using a standard four probes technique. Single crystal samples from different batches with different crystalline quality were characterized. Field and temperature dependent Hall resistance $R_{xy}(B,T)$ and MR were measured using a six probes technique. Angle-dependent MR (ADMR) was measured using a rotation option. The magnetic susceptibility with $B\perp(001)$ and $B\parallel(001)$ were measured on a sample in mass about $m$=23.6 mg.

### III. RESULTS AND DISCUSSION

#### A. Crystal Structure

The single crystals were checked by XRD diffraction data (Fig.1(b)). The strong diffraction peaks can be indexed as (00$l$) consistent with the ICDD-PDF 21-1201. Meanwhile, this result indicates that the exposed surfaces of the grown crystals consist mainly of {001} plane. The single crystals are brittle and can be easily cleaved along (001) plane. On the cleaved fresh crystal surface, no traces of iodine (transport agent in crystal growth) were found and the Ta:Te atomic ratio is almost close to 1:2, as confirmed by EDX analysis (See S1). Fig. 1(a) presents the crystal structure of 1T'-TaTe$_2$ (Space group C2/m (No. 12), Z=6, a=19.31 Å, b=3.651 Å, c=9.377 Å, $\beta$=134.22°)[6]. In TaTe$_2$, each Ta atom is surrounded by six Te atoms, resulting in a distorted TaTe$_2$ octahedron (left inset in Fig. 1(a)) connected via common edges, which favors forming two dimensional corrugated slabs. The double zigzag chains of Ta atoms for both LT-TaTe$_2$ and 1T'-

TaTe$_2$ were shown in the insets of Fig. 2 (a). Below the phase transition temperature, the Ta atoms tend to be clustered accompanied forming two types rhombuses along $b$ axis[8], resulting in the unit cell expanding three times larger than that of 1T'-TaTe$_2$.

## B. Magneto-transport Properties of LT-TaTe$_2$

The in-plane resistivity $\rho_{xx}(0, T)$ (Fig. 2(a)) shows a metallic behavior. The resistivity is 207.5 $\mu\Omega$·cm at 300 K, and 3.2 $\mu\Omega$·cm at 3 K. The residual resistance ratio (RRR) defined as $\mathrm{RRR} = \rho_{xx}(300\,K)/\rho_{xx}(3\,K)$ is an important parameter to assess crystal quality. Here, the maximum RRR is about 65, indicating the high crystalline quality of our samples. The Peierls transition temperature is about 170 K at which a steeply decrease in resistivity can be seen. A thermal hysteresis in resistivity is observed between the cooling and warming cycles with a rate of 5 K/min, consistent with the previously reported results[8,9].

Since TaTe$_2$ is a semimetal[14], both electrons and holes are effective carriers. The positive slope of $R_{xy}(B)$ (See Fig. S2) suggests that the dominant charge carriers in TaTe$_2$ are holes. Moreover, the Hall resistance showing anomalous fluctuation at low temperatures, but becomes relatively smooth at high temperatures. The Hall resistivity is almost linear with magnetic field. That means Hall coefficient $R_H = d\rho_{xy}(B)/dB$ is almost a constant and suggests the net carrier concentration $(n_h - n_e) = 1/(eR_H)$ of LT-TaTe$_2$ should be independent of magnetic field. If a single band model of carriers transfer is employed, the carrier mobility can be derived by $\mu = R_H/\rho_{xx}$. The temperature dependence of carrier concentration $(n_h - n_e)$ and carrier mobility are shown in Fig. 2(b). It is noted the carrier concentration decreases with temperature decreasing. The net carrier density $(n_h - n_e)$ at 3 K is about $8\times10^{20}$ cm$^{-3}$, which are about two orders of magnitude lower than conventional metals. The mobility is deduced to be about 1400 cm$^2$/(V s) at 3 K. The carrier concentration measured on sample with RRR~16 reported by Liu et al[9] is almost three times of our sample with RRR=65 in this work.

Measurements of transverse MRs of sample with RRR=65 versus magnetic field at low temperature are presented in Fig. 3(a). A crossover of MR from a semi-classical weak-field $B^2$ dependence to a linear $B$ dependence is observed when the magnetic field is higher than a critical field $B^*$. A relatively large MR reached 140% when magnetic field increased to 9 T at 3 K. In order to determine the critical field $B^*$, we plot the differential of MR versus field, $d$MR/$dB$, as shown in Fig. 3(b). In the low field range ($B<1$ T at 3 K), $d$MR/$dB$ is proportional to $B$ (as shown by a black guide line in low field region), indicating the semi-classical $B^2$ dependence. But above a critical field $B^*$, $d$MR/$dB$ deviates from the semi-classical behavior and saturates to a much reduced slope. With increasing temperature, the critical field shifts to higher field as shown in Fig. S4. A dependence of the critical field on temperature is shown in Fig. 3(c). Above 100 K, the MR shows a semi-classical $B^2$ dependent behavior below 9 T completely.

Besides, the MR is sample-dependent and larger in higher quality samples (See Fig. 3(d)). This is manifested by the fact that the largest MR of RRR=65 is near six times of that with RRR=20 at 9 T. Unlike the samples with small RRR values, the MR variation with magnetic fields for the high quality samples show a saturation-like behavior, which was also found in the Dirac semimetal $Cd_3As_2$[15]. For samples with RRR less than 30, the MRs is nearly a linear behavior (See Fig. S3 for sample with RRR=20) and does not show any signs of saturation within measured field range.

Anomalous linear MR was also observed in other quasi-2D CDW[16] and spin density wave (SDW)[17] compounds such as 2H-$TaSe_2$, 2H-$NbSe_2$, and iron-based superconductor parent material $BaFe_2As_2$. The linear MR in 2H-$TaSe_2$ and 2H-$NbSe_2$ were explained by the reconstruction of Fermi surface and magnetic breakdown through the CDW gaps[16]. In $BaFe_2As_2$, the linear MR was explained to originate from the existence of Dirac cone states which are the nodes of the SDW gaps at low temperature[17, 18]. Apparently the anomalous linear MR in our samples could not be explained only in terms of classical point of view or by extrinsic reasons[16, 19]. Here, we consider the possible origins of the anomalous linear MR closely related to quantum linear magnetoresistance (QLMR) as proposed by Abrikosov[20-22] to explain transport phenomena observed in layered metals. In the quantum limit, at a specific temperature and field, Landau level splitting between the lowest and the first Landau level is larger than both the Fermi energy $E_F$ and the thermal fluctuations $\kappa_B T$, so that all carriers occupy the lowest Landau level and QLMR ensues. In this situation, the following condition should be satisfied [20]:

$$n_e \ll (eB/\hbar c)^{3/2} \qquad (1)$$

The condition (1) implies that only the lowest Landau level is filled with electrons. For a field $B\sim 10$ T, the required carrier concentration is about $10^{18}$ cm$^{-3}$ to observe QLMR. Whereas the net carrier concentration in our $TaTe_2$ (Fig. 2(b)) is of the order of $10^{20}\sim 10^{21}$ cm$^{-3}$, which is much higher than the carrier concentration limit. This indicates a relatively large Fermi surface presents in $TaTe_2$ compared with the required carrier concentration for a QLMR. Abrikosov[21, 22] suggested that apart from the large Fermi surface which can be treated classically, there should exist a small Fermi pocket with almost linear energy dispersion[23] satisfying the condition (1). The effective mass of the small Fermi pocket should be small. The quantum transport of the small Fermi pocket in Dirac cone state in quantum limit will induce a linear MR.

For energy band with linear dispersion[23], the splitting in energy between the lowest and the first Landau level can be described by $\Delta_1 = |E_{\pm 1} - E_0| = \pm v_F \sqrt{2\hbar eB}$. When the quantum limit is approached in a linear energy band, the critical field $B^*$ should satisfy the equation $B^* = (1/2e\hbar v_F^2)(E_F + k_B T)^2$ [17]. As shown in Fig. 3(d), the critical field deduced from our LT-TaTe$_2$ sample could be well fitted by the equation. The fitting results (See Fig. 3(d)) give the Fermi level $E_F \approx 2.65$ meV, and the Fermi velocity $v_F \approx 3.27 \times 10^4$ m/s. The estimated Fermi velocity and Fermi energy are of the same orders as those observed in BaFe$_2$As$_2$, which has been confirmed that the linear MR is originated from the Dirac cone states as supported by magneto-transport[17] and ARPES[18] studies. Based on the relatively high mobility and the good agreement of the evolution of $B^*$ with temperature[17], we further confirm the inference that the linear MR behavior is originated from the Dirac cone states formed by the Fermi surface reconstruction in LT-TaTe$_2$.

The ADMR of a sample with RRR=63 on the polar angle at azimuth angle $\varphi$=90° as shown in Fig. 4 are analyzed in detail. The geometry of ADMR measurements, polar angle $\theta$ and azimuth angle $\varphi$ are defined in the inset of Fig. 4(a). When $\varphi$=90°, the magnetic field of polar ADMR was always perpendicular to the applied current along b axis. As shown in Fig. 4(b), The polar ADMR at 50 K in 9 T has the maximums (MR=18.7%) when $\theta$=0° and 180°. The MR decreases gradually with angle increasing and reaches the minimum (MR=6.2%) at $\theta$=90° where the MR decreased 66.8% compared to the maximum. The polar ADMR at 50 K in 9 T could be well fitted by a function of |cos$\theta$| as shown in Fig. 4(b), which indicates a quasi-2D Fermi surface[24] existed in LT-TaTe$_2$. The small deviation at lower temperature implies existence of 3D electronic transport in TaTe$_2$.

With decreasing temperature, the MR anisotropy of TaTe$_2$ changes gradually as shown in Fig. 4(b). Similar effect of ADMR change was also found in ADMR at 10 K from low fields to high fields as shown in Fig. 4(c). A strong anisotropy is observed for polar ADMR at 10 K in 9 T. The MR at $\theta$=0° is no longer an extreme point as the case at 50 K. Two minimums were found at $\theta$=11° and 90°. The MR at $\theta$=90° decreased to 70% of that at $\theta$=0°. What is more surprising is two maximums appear at $\theta$=48° and 128°. The MR at $\theta$=128° is even 24% larger than $\theta$=0°. Field dependent MRs at each extreme positions of $\theta$ were measured and shown in Fig. 4(a). The MR shows a saturation tendency with field increasing at $\theta=11°$. At other extreme polar angles, MRs show nearly perfect linear behaviors and do not show any signs of saturation in the measured field range. The strong anisotropy becomes more obvious with the temperature decreasing and the field increasing. At low temperatures and high magnetic fields, the MR contributed from the small Fermi pocket in Dirac cone state would dominate and a linear MR is observed. Thus, we suggest that the strong anisotropy of MR should be induced by the specific anisotropy of the small Fermi pocket that contributes to the linear MR.

### C. Magnetic Anisotropy and de Hass-van Alphen Oscillations of LT-TaTe$_2$

To further clarify the topology of Fermi surface we studied the magnetic properties and de Hass-van Alphen (dHvA) oscillations of TaTe$_2$. The magnetization at $B\perp(001)$ as a function of temperature is shown in Fig. 5(a). A competition between paramagnetism and diamagnetism is observed. At low temperatures, it is paramagnetic and its susceptibility decreases with rising temperature, at a certain temperature it changes into diamagnetic. There's a steep drop of susceptibility at $T_{CDW}$ as shown in Fig. 5(b) and Fig. 5(d). The magnetization at $B\parallel(001)$ (Fig.5 (c)) does not show so strong paramagnetism as in the case with $B\perp(001)$, and it is diamagnetic over the whole measured temperature range. The susceptibility of TaTe$_2$ at high temperature and high magnetic field shows strong fluctuation, which might be related to the fluctuation of CDW in TaTe$_2$[1]. It should be noted that the small peaks observed around 50 K in Fig. 5(b) and Fig. 5(d) are possibly due to the adsorbed oxygen[25].

When temperature is below or equal to 10 K, dHvA oscillations are observed both on the magnetization for $B\perp(001)$ and $B\parallel(001)$ as shown in Fig. 5(a) and (c). We extracted the quantum oscillations by subtracting the backgrounds, as shown in Fig. 6(a) and (b). Numerical Fourier transformation was used to extract the frequencies of the oscillations (see insets in Fig. 6(a) and Fig. 6(b)). The size of Fermi surface cross-sections is related to Onsager relationship[26] by $F=(\hbar c/2\pi e)A_F$, where $A_F$ is the Fermi surface cross-section area, $\hbar$ and $c$ are reduced Planck constant and light velocity, respectively. When a magnetic field was perpendicular to (001), the Fourier transformation spectrum (inset in Fig. 5(a)) of the oscillation at 3K exhibits three frequencies at 24 T, 42 T, and 68 T which are labeled by α, β, and γ respectively. And the corresponding $A_F$ of α, β, and γ are $2.28\times10^{-3}$ Å$^{-2}$, $4.03\times10^{-3}$ Å$^{-2}$, and $6.38\times10^{-3}$ Å$^{-2}$ respectively. When the field was paralleled to (001), two frequencies were found at 23 T and 147 T which are labeled as δ and ε respectively as shown in Fig. 6(b). The area of the relatively large cross-section ε is about $1.38\times10^{-2}$ Å$^{-2}$ which occupies up 1.85% of the projection area of the first Brillouin zone into the (001) plane in reciprocal vector space of 1T'-TaTe$_2$, and the area of the smallest cross-section δ is about $2.17\times10^{-3}$ Å$^{-2}$, which only takes up 0.28% of the projected Brillouin zone area.

The determined five sets of Fermi surface cross-sections should not be participating in linear MR phenomena. This is due to the carriers in contributition to the linear MR should only occupy the lowest Landau level. In the case, there is no sign of dHvA oscillation could be observed. So the observed dHvA oscillations confirm the existence of the relatively large Fermi surface as stated in linear MR part. In fact, the determined five sets of Fermi surface cross-sections are very small compared with normal metals. The consequence should result from the complicated band fold and the Fermi surface reconstruction below the CDW transition temperature. Yet, no signal of Shubinikov-de Haas (SdH) oscillation has ever been observed in resistivity measurement. That is due to the relatively strong linear MR signal hide the MR contributed from the carriers in the relative large Fermi surfaces.

## IV. CONCLUSION

In conclusion, anisotropic magneto-transport and magnetic properties of LT-TaTe$_2$ were studied systematically based on high quality single crystals of TaTe$_2$ with carrier mobility over 1400 cm$^2$/(V s) at low temperature. Observed anisotropic QLMR at low temperatures and high fields in LT-TaTe$_2$ indicate a quasi-2D characteristic of main part of Fermi surface and existence of an anisotropic small Fermi pocket in Dirac cone state. The well fitted critical field $B^*$ versus temperature strongly supports the inference of there existed a small Fermi pocket in Dirac cone state. The magnetic measurements reveal that there is a relatively strong paramagnetic signal when $B\perp(001)$, but a very weak paramagnetic signal when $B\|(001)$ at low temperature; and five sets of small Fermi surface cross-sections are derived from anisotropic dHvA oscillations, consistent with existed relatively large Fermi pockets as discussed in QLMR. Our results and analysis provide new information about the Fermi surface of LT-TaTe$_2$ and show the possibility of Dirac cone states in this quasi-2D CDW compound.

**Acknowledgements**

The authors would like to thank Mr. J. Zhang of Institute of Physics, Chinese Academy of Sciences, for helpful discussion. This work is financially supported by the Key Research Program of Frontier Sciences, CAS, Grant No. QYZDJ-SSW-SLH013, the National Natural Science Foundation of China (Grant Nos. 91422303, 51532010, 51472265, and 51272279) and the National Key Research and Development Program of China (2016YFA0300600).

# Figures.

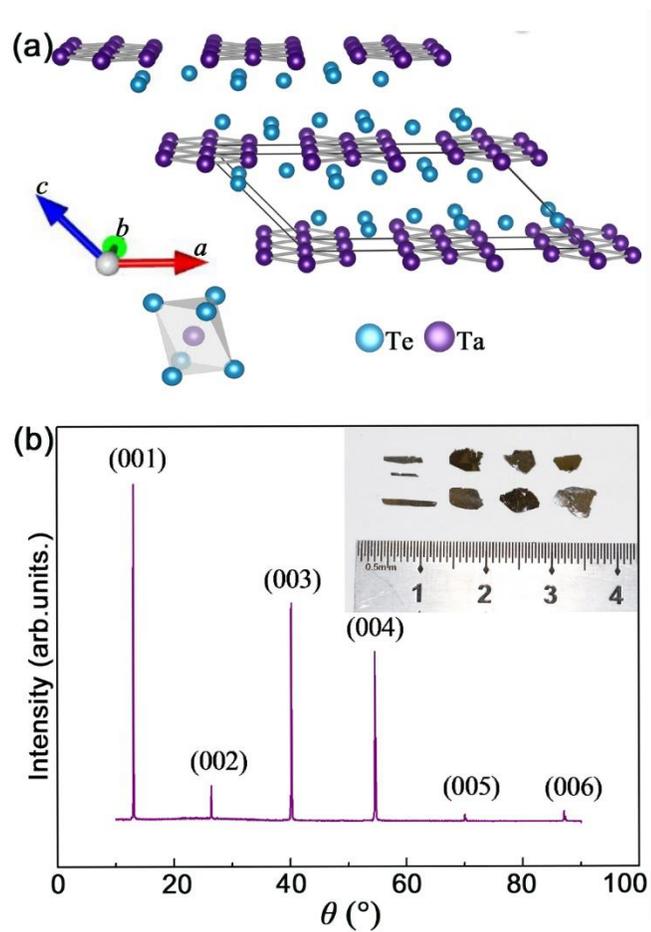

FIG. 1. (a) The crystal structure of 1T'-TaTe$_2$[5]. The solid lines indicated the unit cell in 1T'-TaTe$_2$. (b) X-ray diffraction pattern of a TaTe$_2$ single crystal measured at room temperature. Inset: an optical image of typical single crystal samples from different batches.

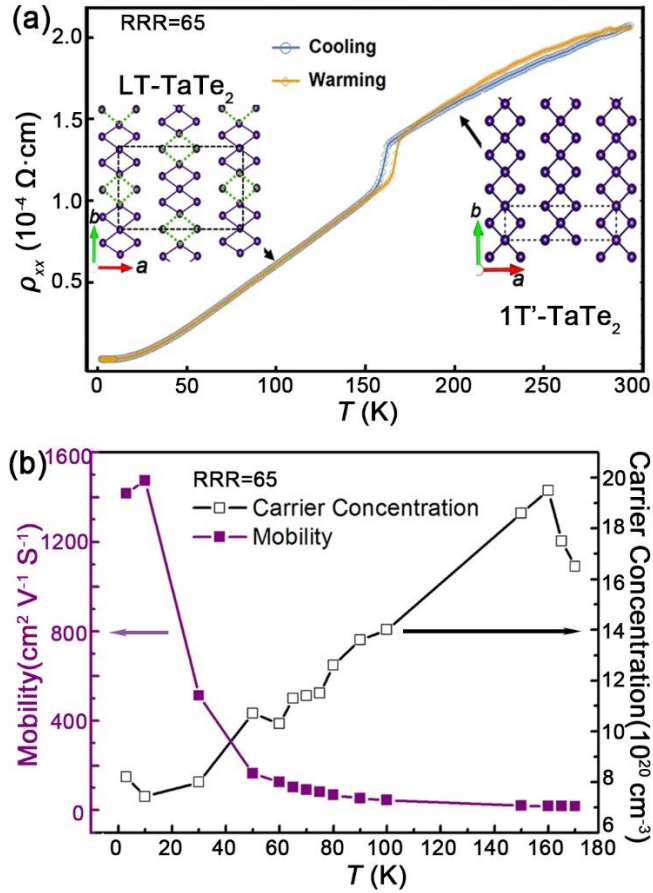

FIG. 2. Transport properties of TaTe$_2$. (a) Zero-field in-plane resistivity of TaTe$_2$, I∥b, the insets are schematic diagrams to show the arrangements of Ta atoms in Ta atoms plane of 1T'-TaTe$_2$ and LT-TaTe$_2$[6, 8]. The dashed rectangles demonstrate the unit cells in 1T'-TaTe$_2$ and LT-TaTe$_2$. The two types rhombuses were sketched by solid purple lines and dotted green lines respectively. (b) Net carrier concentration ($n_h$-$n_e$) and mobility of the sample with RRR=65 at low temperatures.

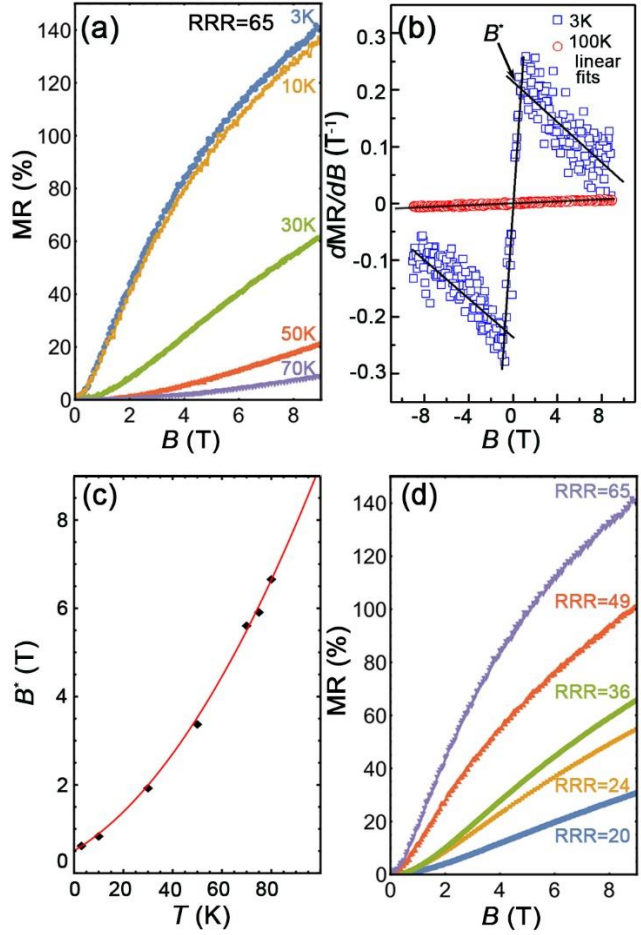

FIG. 3. (a) The field dependent MR of the sample with RRR=65 at low temperatures (3 K~70 K). (b) Dependence of MR differentiation on magnetic field at 3 and 100 K for sample with RRR=65. A critical field $B^*$ is defined as the intersection between the two lines that are just linear fitting of $d$MR/$dB$ in low field semi-classical regime and high field quantum regime. (c) Dependence of critical field $B^*$ on temperature as marked by black squares. The red solid curve is a fit with $B^* = (1/2e\hbar v_F^2)(E_F + k_B T)^2$. (d) Dependence of MRs on magnetic field for samples with different RRR values at 3 K.

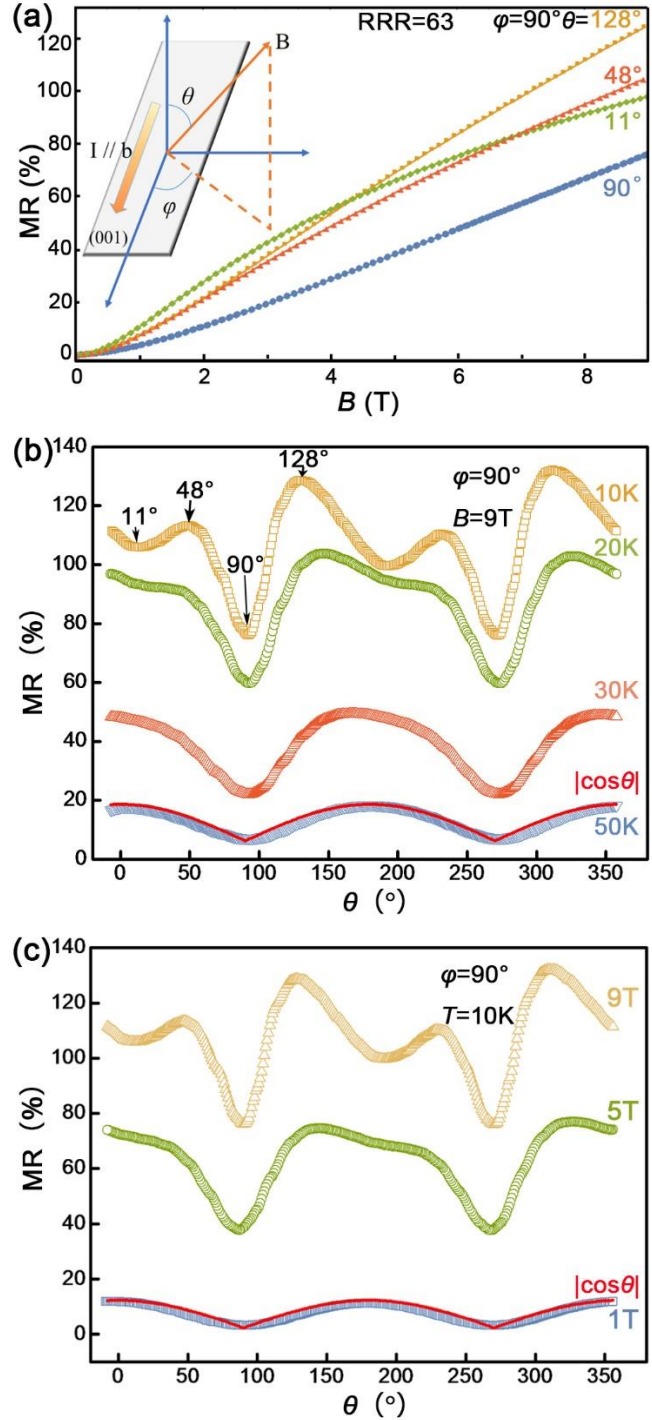

FIG. 4. Anisotropic MRs (a) Field dependent MR at four extreme azimuth angle positions (11°, 48°, 90°, and 128°). Inset: the geometry of angle dependent MR measurements. (b) Polar angle $\theta$ ($\varphi$=90°) dependent MR of sample with RRR=63 at different temperatures in 9 T. The fitting curve fitted by the function of $|\cos\theta|$ is shown by the red curve. The maximums of MR at 3 K are indicated by the arrows. (c) Polar angle $\theta$ ($\varphi$=90°) dependent MR of sample with RRR=63 at 1 T, 5 T, and 9 T at 10 K. The fitting curve fitted by the function of $|\cos\theta|$

is shown by the red curve.

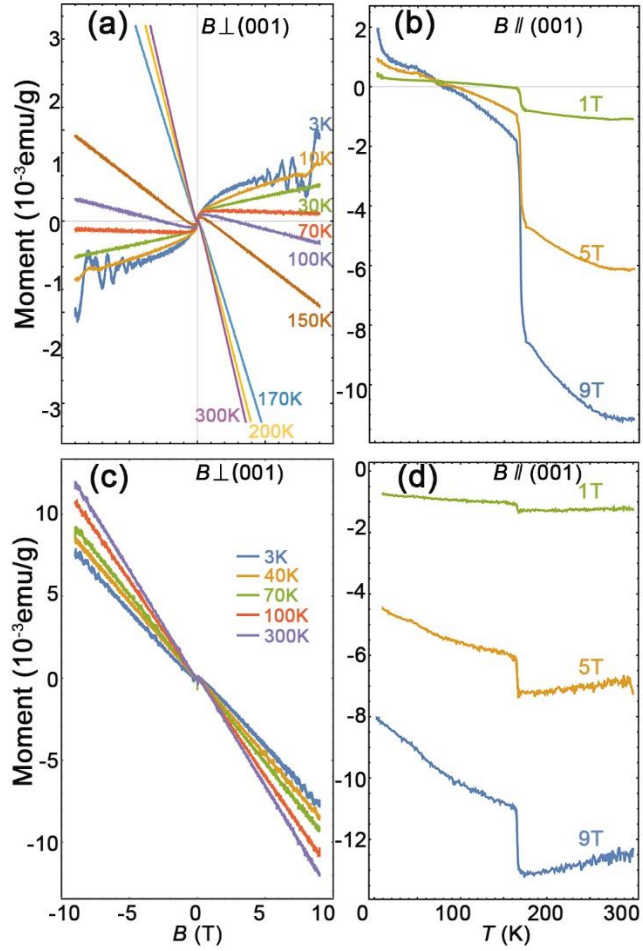

FIG. 5. Magnetic properties of TaTe$_2$. (a) Magnetization of $B\perp(001)$. (b) Temperature dependent susceptibility of $B\perp(001)$ in different fields. (c) Magnetization of $B\parallel(001)$. (d) Temperature dependent susceptibility of $B\parallel(001)$ in different fields.

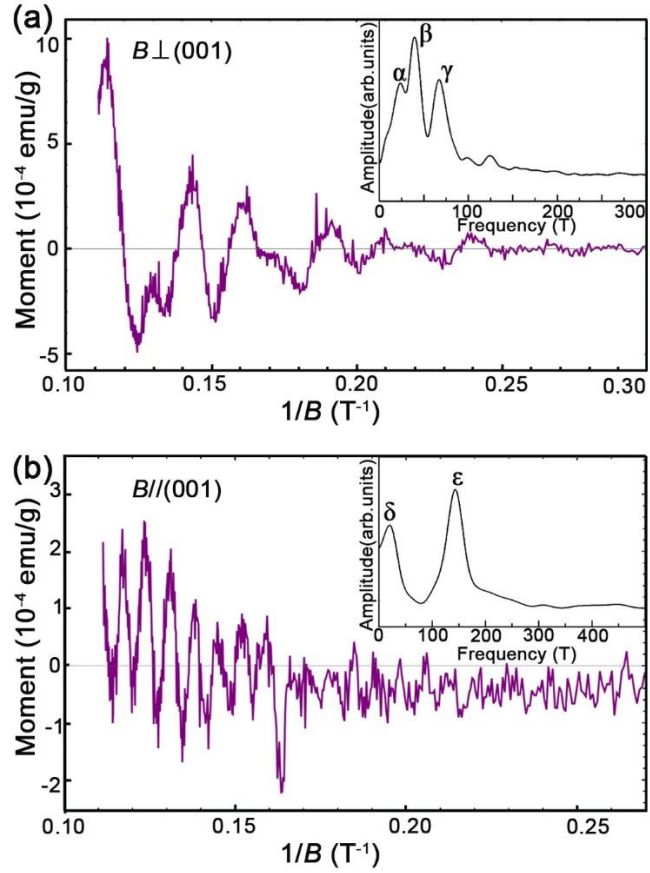

FIG. 6. dHvA oscillations in LT-TaTe$_2$ (a) dHvA oscillatory components of $B\perp(001)$ at 3 K. Inset: Fourier transform of the data in (a), showing three frequencies, labelled as α, β, and γ respectively. (b) dHvA oscillatory components of $B\parallel(001)$ at 3K. Inset: Fourier transform of the data in (b), showing two frequencies, labelled as δ and ε respectively.